\documentclass[10pt,letterpaper]{article}
\usepackage[top=0.85in,left=2.75in,footskip=0.75in]{geometry}

\usepackage{amsmath,amssymb}

\usepackage{changepage}

\usepackage[utf8x]{inputenc}

\usepackage{textcomp,marvosym}

\usepackage{cite}

\usepackage{nameref,hyperref}

\usepackage[right]{lineno}

\usepackage{microtype}
\DisableLigatures[f]{encoding = *, family = * }

\usepackage[table]{xcolor}

\usepackage{array}

\newcolumntype{+}{!{\vrule width 2pt}}

\newlength\savedwidth

\newcommand\thickhline{\noalign{\global\savedwidth\arrayrulewidth\global\arrayrulewidth 2pt}%
\hline
\noalign{\global\arrayrulewidth\savedwidth}}

\raggedright
\usepackage[aboveskip=1pt,labelfont=bf,labelsep=period,justification=raggedright,singlelinecheck=off]{caption}

\makeatletter
\renewcommand{\@biblabel}[1]{\quad#1.}
\makeatother

\date{}

\usepackage{lastpage,fancyhdr,graphicx}
\usepackage{epstopdf}
\fancyheadoffset[L]{2.25in}
\fancyfootoffset[L]{2.25in}
\usepackage{todonotes}

\renewcommand{\H}{\mathsf{H}}
\newcommand{\betti}{\mathsf{b}}
\newcommand{\dnn}{d}

\newcommand{\ms}{MS}
\newcommand{\sm}{SM}

\newcommand{\prob}[1]{P_{#1}}
\newcommand{\step}{\ell}
\newcommand{\turn}{\theta}

\newcommand{\turnspread}{\rho}

\renewcommand{\exp}[1]{\mathrm{e}^{#1}}

\begin{document}
\vspace*{0.2in}

\begin{flushleft}
{\Large
\textbf\newline{Assessing biological models using topological data analysis} 
}
\newline
\\
M. Ulmer\textsuperscript{1},
Lori Ziegelmeier\textsuperscript{1,$\ast$},
Chad M. Topaz\textsuperscript{2}
\\
\bigskip
\textbf{1} Department of Mathematics, Statistics, and Computer Science, Macalester College, Saint Paul, Minnesota, United States of America
\\
\textbf{2} Department of Mathematics and Statistics, Williams College, Williamstown, Massachusetts, United States of America
\\
\bigskip

$\ast$ E-mail: \texttt{lziegel1@macalester.edu}

\end{flushleft}

\section*{Abstract}
We use topological data analysis as a tool to analyze the fit of mathematical models to experimental data. This study is built on data obtained from motion tracking groups of aphids in [Nilsen et al., PLOS One, 2013] and two random walk models that were proposed to describe the data. One model incorporates social interactions between the insects, and the second model is a control model that excludes these interactions. We compare data from each model to data from experiment by performing statistical tests based on three different sets of measures. First, we use time series of order parameters commonly used in collective motion studies. These order parameters measure the overall polarization and angular momentum of the group, and do not rely on \emph{a priori} knowledge of the models that produced the data. Second, we use order parameter time series that do rely on \emph{a priori} knowledge, namely average distance to nearest neighbor and percentage of aphids moving. Third, we use computational persistent homology to calculate topological signatures of the data. Analysis of the \emph{a priori} order parameters indicates that the interactive model better describes the experimental data than the control model does. The topological approach performs as well as these \emph{a priori} order parameters and better than the other order parameters, suggesting the utility of the topological approach in the absence of specific knowledge of mechanisms underlying the data.


\section*{Introduction}

Given data potentially described by various mathematical models, how might one choose between those models? In the context of experiments on social insects, we use topological data analysis (TDA) to inform this choice, and compare the topological approach to more traditional methods. Our investigation complements the rich literature on biological model selection \cite{BurAnd2003,JohOml2004}.

In order to frame our investigation, we recapitulate key points of the biological experiments that form the basis of our work, namely those of \cite{NilPaiWar2013}. In these experiments, pea aphids, \emph{Acyrthosiphon pisum}, are filmed from above while they move in a circular experimental arena. For each experimental trial, motion tracking software extracts time series of aphids' positions. Aphids move intermittently, and so from this raw data, one may also estimate the motion state -- moving or stopped -- of each aphid in each movie frame. Finally, if an aphid is moving, one may estimate its direction of motion, also known as its heading.

The authors of \cite{NilPaiWar2013} propose two mathematical models. Both models describe transitions between an aphid's motion state with a random process (a biased coin flip). For moving aphids, both models describe the movement process as an unbiased, correlated random walk (described in more detail later). In the first model, referred to as the interactive model, for each aphid and at every moment in time, the motion state transition probabilities and the parameters of the random walk all depend on the distance to an aphid's nearest neighbor. This  dependence is a simple way to account for the (potential) effect of nearby aphids due to social interactions. The parameters in the model are fit to the data using a simple optimization procedure. The second model, referred to as the control model, is similar to the first, except that the transition probabilities and random walk parameters are all taken to be constant, and hence, there are no social effects incorporated.

To assess the models, the authors of \cite{NilPaiWar2013} choose several measures extracted from all of the raw data pooled over the experimental trials: the cumulative distribution of distance to an aphid's nearest neighbor; the cumulative distribution of angle between an aphid's direction of movement and the location of its nearest neighbor;  and finally, the cumulative distribution  of the proportion of aphids moving in each frame.  They  then make comparisons between these measures for the experimental data vs. the interactive model and the experimental data vs. the control model. For each of these comparisons, they use three measures of the distance between distributions: the difference between the median, the Kolmogorov-Smirnov distance, and the Kullback-Leibler divergence. For the distributions of distance to nearest neighbor and proportion of aphids moving, the distributions from the interactive model are closer to the  experimental distributions than are those of the control model. For the distributions of angle to nearest neighbor, no significant difference is seen.

With this prior work now reviewed, we ask the question: what if the authors had chosen other quantities to compare? Distance to nearest neighbor, angle to nearest neighbor, and proportion of aphids moving are properties thought by the authors to be potentially important. However, there are many other aggregate measures that one might choose to select between models. Additionally, we note that if the authors had chosen only to examine angle to nearest neighbor, they might have concluded that the social interaction model and the control model are indistinguishable.

We will use tools from topological data analysis -- specifically, persistent homology -- to choose between models.  Loosely speaking, topological data analysis  looks at the shape of data, which is arguably a more neutral property than biological or physical measures that are derived from the data and are selected by an experimentalist or modeler. 

As merely one example of the power of topological data analysis, we mention \cite{TopZieHal2015}, which performs topological data analysis on numerical simulations of two influential models of biological aggregation \cite{VicCziBen1995,DOrChuBer2006}. In this study, biological group properties that are traditionally used in the literature -- polarization of group motion, angular momentum, and so forth -- are calculated for simulations conducted in different parameter regimes. The authors then compare these traditional measures with topological signatures of the data calculated using persistent homology. The topological signatures reveal important dynamical differences between simulations for which the traditional measures are similar, and conversely, detect topological similarity between simulations for which the traditional measures appear somewhat different. More broadly, topological data analysis is gaining ground as a methodology for studying biology, and has recently been applied to brain architecture \cite{GiuGhiBas2016}, cancer genetics \cite{NicLevCar2011}, epidemic networks \cite{TayKliHar2015} and much more.

In our investigation, for each of nine experimental trials on aphid motion, we compare data sets from experiment with those generated by the interactive and control models. We carry out these comparisons using three sets of measures. First, we use time series of order parameters frequently used in physics to assess collective motion, namely polarization, angular momentum, and absolute angular momentum. Then, we use a second set of order parameters, namely average nearest neighbor distance and percent of aphids moving. Since the control model does not incorporate social interactions and the interactive model does, one might naturally expect that this second group of order parameters should strongly reflect the model inputs. Third, we compare data using topological signatures called crockers, defined in \cite{TopZieHal2015}.

Using these three sets of measures, we perform statistical tests to asses whether the interactive model or the control model better describes the experimental data. Analysis of the second set of order parameters, which use \emph{a priori} knowledge of the model, indicates that the interactive model better describes the experimental data. The topological approach performs as well as these order parameters and better than the first set of order parameters, suggesting the utility of the topological approach in the absence of specific knowledge of mechanisms underlying the data.

The rest of this paper is organized as follows.  First, we provide additional detail on the experiments and models of \cite{NilPaiWar2013}, as well as on our computational implementation of the models. Then, we review key ideas from topological data analysis, with a focus on a tool known as computational persistent homology. Finally, to carry out the core of our study, we calculate order parameter time series and topological signatures, perform statistical tests on them, and compare the various approaches.

\section*{Models of aphid motion} \label{sec:models}

\begin{figure}[t!]
\centerline{\includegraphics[width=0.8\textwidth]{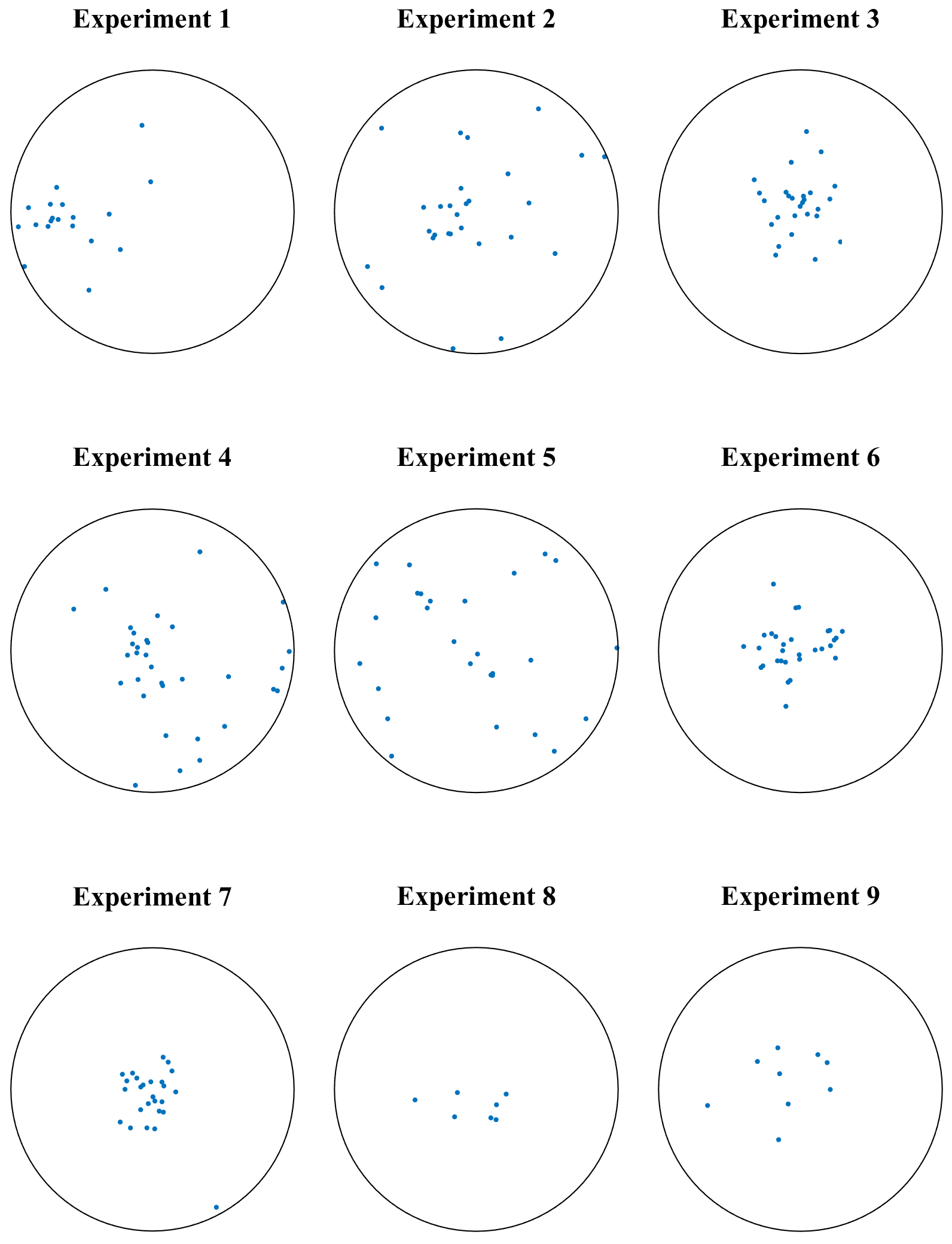}}
\caption{{\bf Initial conditions of experimental trials from \cite{NilPaiWar2013}.}
In nine separate experimental trials, pea aphids were placed in a featureless circular arena $40\ cm$ in diameter and filmed from above. Here we  show the initial positions of the aphids for each trial, as obtained from image processing and motion-tracking software. Trials one through nine are composed of, respectively, 19, 28, 27, 33, 27, 30, 26, 7, and 9 aphids. See Fig \ref{fig2} for an example of how an initial condition evolves over time. Data were obtained from the authors of \cite{NilPaiWar2013}.}
\label{fig1}
\end{figure}

\begin{figure}[t!]
\centerline{\includegraphics[width=\textwidth]{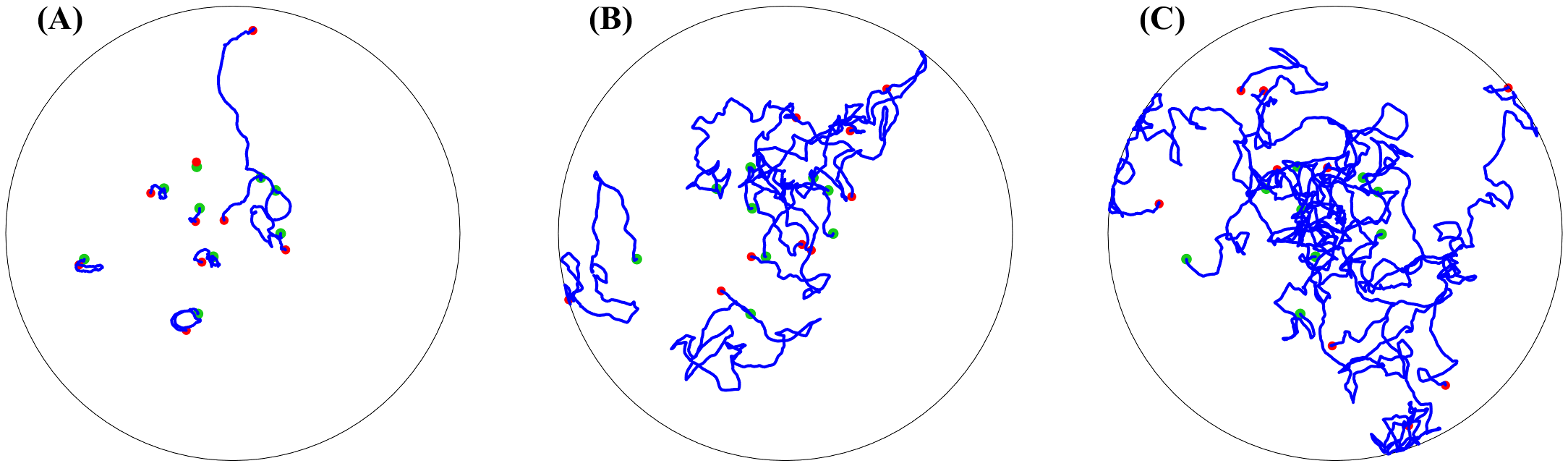}}
\caption{{\bf Example aphid trajectories.}
Green points represent aphids' initial positions, red points their final positions, and the blue curves are their paths over all other frames. The enclosing circle marks the boundary of the arena. (A) Movement of aphids over the first 900 frames of Experiment 9. (B) Trajectories from a realization of the interactive model with initial configuration from Experiment 9. (C) Trajectories from a realization of the control model with initial configuration from Experiment 9.  }
\label{fig2}
\end{figure}

We reprise some details of the experiments and modeling in \cite{NilPaiWar2013}. During nine experimental trials, groups of 7 to 33 aphids were filmed for 45 minutes walking in a circular arena $40\ cm$ in diameter. Fig \ref{fig1} shows the initial state of the aphids in each trial. Fig \ref{fig2}(A) shows trajectories of the aphids in Experiment 9, obtained via motion-tracking software. 

The model of \cite{NilPaiWar2013} classifies each aphid according to two motion states, moving or stationary, during each $0.5\ s$ video frame. Transitions between these states are probabilistic, with the probability of transition for each aphid depending on the distance $\dnn$ to that aphid's nearest neighbor. Moving aphids obey an unbiased correlated random walk. According to this model, over the course of one frame, each moving aphid takes a step of a certain length that depends on $\dnn$. Then, it turns at a random angle drawn from a distribution that is centered around zero. The spread of the distribution also depends on $\dnn$.

As modeled in \cite{NilPaiWar2013}, the probability $\prob{\ms}(\dnn)$ of a moving aphid stopping in the next frame is 
\begin{equation}
\prob{\ms}(\dnn) = \prob{\ms}^\infty + \left(\prob{\ms}^0 - \prob{\ms}^\infty\right) \exp{-\dnn/d_{MS}},
\label{eq:pmsfunctionalform}
\end{equation}
with $\prob{\ms}^\infty \approx 0.1280$, $\prob{\ms}^0 \approx 0.5508$, and $d_{MS} \approx 0.0134\ m$. The probability $\prob{\sm}(\dnn)$ of a stationary aphid moving in the next frame is
\begin{equation}
\label{eq:psmfunctionalform}
\prob{\sm}(\dnn)= \prob{\sm}^0 \exp{-\dnn/d_{SM}} + \prob{\sm}^\infty \frac{\dnn}{\dnn+\Delta_{SM}},
\end{equation}
with $\prob{\sm}^0 \approx 0.1587$, $\prob{\sm}^\infty \approx 0.3552$, $d_{SM} \approx 0.0079\ m$, and $\Delta_{SM} \approx 0.0739\ m$.

For moving aphids, the step length $\step(\dnn)$ is
\begin{equation}
\step(\dnn) = \step^\infty + \left(\step^0 - \step^\infty\right) \exp{-\dnn/d_{\step}},
\label{eq:stepfunctionalform}
\end{equation}
with $\step^\infty \approx 0.0013\ m$, $\step^0 \approx 0.0003\ m$, and $d_{\step} \approx 0.0074\ m$. The turning angle $\turn$ is drawn from a wrapped Cauchy distribution centered at zero,
\begin{equation}
\label{eq:wrappedcauchy}
f(\turn)=\frac{1}{2\pi}\frac{1-\turnspread^{2}}{1+\turnspread^{2}-2\turnspread\cos \theta}.
\end{equation}
The parameter $0<\turnspread<1$ controls the spread of the distribution, with small values producing wider distributions and values closer to one producing strongly peaked ones. The spread parameter is
\begin{equation}
\turnspread(\dnn) = \turnspread^\infty + \left(\turnspread^0 - \turnspread^\infty\right) \exp{-\dnn/d_{\turnspread}},
\label{eq:turnspreadfunctionalform}
\end{equation}
with $\turnspread^\infty \approx 0.9013$, $\turnspread^0 \approx 0.1387$, and $d_{\turnspread} \approx 0.0044\ m$.

Descriptions in \cite{NilPaiWar2013} provide biological interpretations of the functional forms above, the roles of the parameters therein, and the procedure for fitting the models to data in order to determine parameter values. For our purposes, the key point is that all four quantities $\prob{\ms}$, $\prob{\sm}$, $\step$, and $\turnspread$ depend on distance to nearest neighbor, $\dnn$. Through this dependence, social interactions enter into the model, and hence we refer to this model as the \emph{interactive model}.

On the other hand, a \emph{control model} from \cite{NilPaiWar2013} ignores the effects of social interactions. The model is identical to the model above, but with $\dnn \to \infty$; that is to say, the model assumes aphids do not interact, so that $\prob{\ms}$, $\prob{\sm}$, $\step$, and $\turnspread$ are constant.

Later, we will present results on numerical simulations of the interactive model and the control model described above. Simulations are seeded with the initial conditions extracted from the motion-tracking experiments. In simulating the models, it is necessary to implement a boundary condition. We use a simple rebound-type condition; aphids passing through the boundary of the experimental arena on a given time step are reflected back into the arena. 

Fig \ref{fig2} displays trajectories of sample simulations of both the interactive and control models along with the corresponding experimental data from Experiment 9. We observe that in the experimental data some aphids were largely stationary, while others moved more around the arena. This is in contrast to both of the sample simulations, where aphid motion is much more pronounced, particularly in the control model. 

\section*{Topological Data Analysis}

Later, we will analyze our data with tools from topological data analysis (TDA). For now, we explain these methods. We provide a brief and nontechnical review of the main ideas followed by a slightly more technical explanation for the mathematically-inclined reader.

Topology is the branch of mathematics concerned with studying properties of objects that are preserved when that object is stretched, compressed, or bent (but not torn or glued). For example, a sphere and a cube are topologically equivalent because one can be bent into the other; similarly for a hollow circle and a hollow triangle. A hollow circle and a filled-in circle are not topologically equivalent because the former has a hole in the middle while the latter does not. To take a topological approach to data, we imagine that data points gathered from simulation or experiment represent some object. For instance, suppose we wish to analyze the position data of aphids from a single frame of experiment. Mathematics provides a way to create an object by taking our data points and connecting them with line segments, triangles, tetrahedrons, and other simple pieces. Then, we quantify the shape of this object by calculating the number of holes of different dimensions using topology. These quantities are called \emph{Betti numbers}, and we say that they specify the object's \emph{homology}. However, the particular object built from the data depends on a scale (distance) parameter called the \emph{proximity parameter} or \emph{filtration value}. We repeat the aforementioned process for many different values of this parameter and examine how homology changes with respect to this parameter. This is called calculating the \emph{persistent homology}, and the output of this computation reveals a topological signature of our data. This entire process has been focused on data from a single time step. We can repeat the entire process for every time step, thereby producing a topological signature of our data over different values of proximity parameter and time.

More technically, our data can be viewed as a (potentially noisy) sampling from an underlying manifold or topological space. If we augment our data with a notion of distance between points then we can use persistent homology to understand the topological structure of our data at a variety of scales; see, \emph{e.g.}, \cite{ZomCar2005,Ghr2008,EdeHar2008,Car2009}.

\emph{Homology} is a tool from algebraic topology that measures the features of a topological space \cite{Hat2002,Mun1993,Cro2006}. Loosely speaking, homology measures the quantity of holes with boundaries of different dimensions $k$, and this quantity is encoded in the Betti number $\betti_k$. The number of connected components (\emph{i.e.}, clusters) corresponds to $\betti_0$, the number of topological circles (\emph{i.e.}, loops) corresponds to $\betti_1$, the number of trapped volumes (\emph{i.e.}, voids) corresponds to $\betti_2$, and so on.

One method of computing homology is via a simplicial complex, a collection of simple pieces such as vertices, edges, filled triangles, solid tetrahedra, and so on, called $k$-simplices for $k=0,1,2,3,\ldots$, respectively. While there are a variety of approaches to create a simplicial complex, the \emph{Vietoris-Rips} complex is a common choice as it is computationally tractable \cite{Ghr2008}. This simplicial complex is created by identifying data points as vertices and forming a $k$-simplex whenever $k+1$ points are pairwise within some distance $\epsilon$, called the \emph{proximity parameter}. Therefore, the structure of the simplicial complex is highly dependent on this choice of parameter. Homology determines the $k$-cycles (in the case of a 1-cycle, this is analogous to a cycle within a graph) that are not boundaries of $k+1$-simplices by splitting all $k$-cycles into equivalence classes, one for each distinct hole in the simplicial complex. Algebraically, the $k$th homology $\H_k$ can be viewed as a quotient of vector spaces with dimension equal to the Betti number $\betti_k$. The Betti number $\betti_0(\epsilon)$ corresponds to the number of connected components at scale $\epsilon$ and $\betti_1(\epsilon)$ corresponds to the number of cycles at scale $\epsilon$. 

\begin{figure}[t!]
\centerline{\includegraphics[width=1\textwidth]{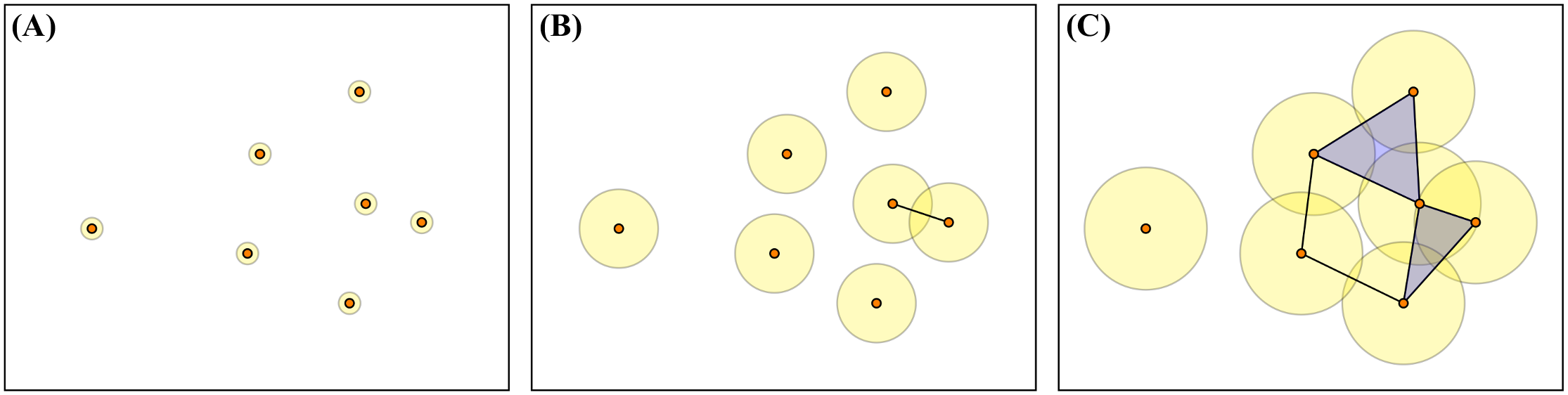}}
\caption{{\bf Simplicial complex forming over increasing proximity parameter $\epsilon$.} Points are joined into simplices as $\epsilon$ increases. The orange points represent 0-simplices, the data points on which the complex is built. These are connected by 1-simplices, represented by lines between the points, and by 2-simplices, represented by blue triangles. The yellow circles have radii of $\epsilon/2$, with intersections indicating where simplices form; however, the circles are for visualization only, and are not part of the simplicial complex.  In (A), where $\epsilon$ is close to zero, each point is a discrete component. In (B), two points have come within $\epsilon$ and are joined by a 1-simplex. In (C), 2-simplices have formed where there are sets of three points pairwise within $\epsilon$. Note that the upper 2-simplex connects three points whose $\epsilon/2$-balls have no common intersection, a phenomenon possible in the Vietoris-Rips complex. At this scale a 1-dimensional hole (cycle) has formed among four of the data points.}
\label{fig3}
\end{figure}

To eliminate the need to choose a single proximity parameter $\epsilon$, \emph{persistent homology} exploits the fact that as $\epsilon$ grows, so do the Vietoris-Rips complexes, giving an inclusion of simplicial complexes for smaller $\epsilon$ into those for larger values. Fig \ref{fig3} shows the formation of a Vietoris-Rips complex over increasing $\epsilon$ for a fixed point cloud of data. Persistent homology tracks homology classes as the proximity parameter is varied, recording the interval over which a $k$th order class (hole) first appears and when it no longer remains. These intervals represent the topology of a single-parameter family of spaces, a \emph{filtration}, and are complete discrete invariants, meaning they capture all of the topological information in this family. For details regarding persistent homology and its computation, see \cite{Car2009,EdeLetZom2000,EdeHar2008}. 

\begin{figure}[t!]
\centerline{\includegraphics[width=0.8\textwidth]{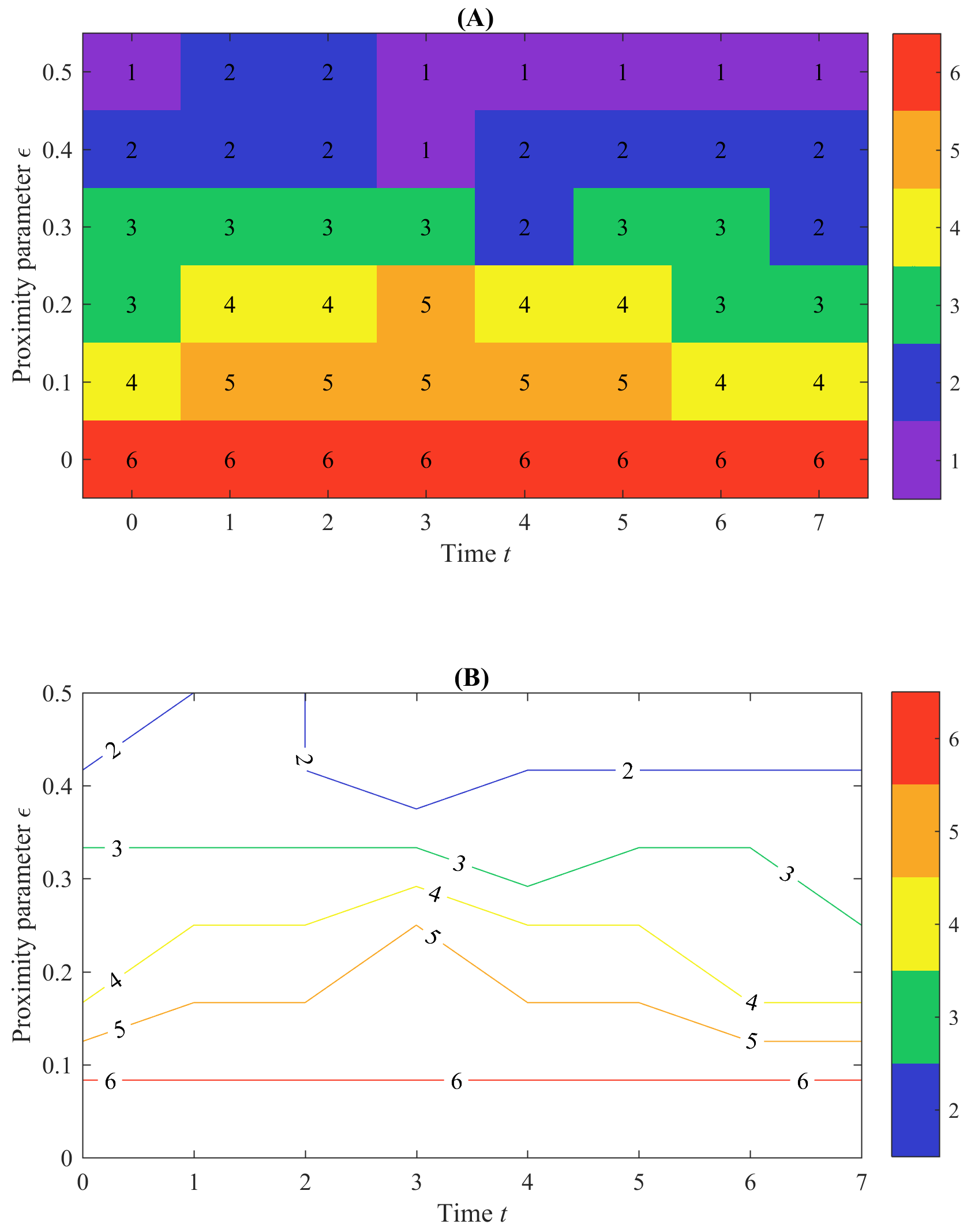}}
\caption{{\bf Example construction of a crocker plot.}
Crocker plots represent the value of a Betti number as a function of simulation/experiment time $t$ and topological proximity parameter $\epsilon$. (A) Fictitious data used to generate an example crocker of Betti number $\betti_0(t,\epsilon)$. The $y$-axis corresponds to discrete values of the proximity parameter $\epsilon$ and the $x$-axis corresponds to discrete time steps. Any entry in the matrix represents the value of $\betti_0$ at a specific time and proximity scale. We apply the same color to all entries having the same value of $\betti_0$. (B)~Crocker plot corresponding to the data in (A). The contours delineate regions of constant $\betti_0$. The region where $\betti_0 = n$ falls between the contours labeled $n$  and $n+1$. }
\label{fig4}
\end{figure}

In a variety of applications, including the analysis of biological aggregations, one may wish to analyze the topological structure of data as another parameter is varied in addition to proximity, for example, density or time. Unfortunately, there is no complete discrete invariant for multiple parameters similar to the interval representation for single-parameter persistent homology \cite{CarZom2009}. However, Betti numbers can still be computed as a function of multiple inputs. In \cite{TopZieHal2015}, the authors propose the Contour Realization Of Computed $k$-dimensional hole Evolution in the Rips complex (crocker) as a means for visualizing the change in Betti number over two parameters. A function over two parameters can be conveniently visualized as a contour plot. Fig \ref{fig4} is a toy example of a crocker plot which displays the Betti number, $\betti_k(t,\epsilon)$ as a function of the two inputs of simulation time $t\geq 0$ and proximity parameter $\epsilon \geq 0$, built by taking a discrete sampling of each parameter and displaying $\betti_k(t,\epsilon)$ first in matrix form in Fig \ref{fig4}(A) and as a contour plot in Fig \ref{fig4}(B). Large regions in the contour diagram represent topological features that persist over both the scale $\epsilon$ and simulation time $t$. Alternative approaches to crocker plots are discussed as the dimension function of a 2-D persistence module in \cite{LesWri2015} or pseudo-multidimensional persistence in \cite{XiaWei2015}.

It should be noted that for a fixed time, the corresponding vertical cross section represents a filtration through the proximity parameter $\epsilon$, sometimes referred to as a \emph{Betti curve} \cite{GiuPasCur2015}. In Fig \ref{fig4}, $t=0$ represents the fixed point cloud from Fig \ref{fig3} above. Cross sections corresponding to a fixed $\epsilon$ are not a filtration as there is not necessarily inclusion in the simplicial complex from one time step to the next. However, as the aggregation models we study evolve smoothly in time, we expect the topological features to do so as well.

Crocker plots represent a matrix of data, and hence can be vectorized, resulting in a compact topological summary with flexible uses. First, since crockers can be viewed as vectors, they can be used in the application of myriad machine learning tools. For instance, one could use machine learning to predict model outcomes based on parameters.

Second, with this representation of the data, distinct homological dimensions may be concatenated together. For instance, the $\betti_{0(pos)}$ and $\betti_{1(pos)}$ crockers could be concatenated to simultaneously compare connected components and topological loops of a simulation. In some applications, \emph{e.g.}, \cite{AdaEmeKir2017}, concatenating homological dimensions yields superior results than analyzing dimensions separately.

Third, crockers may be computed for any multi-parameter family of persistence modules and not just restricted to proximity and time, as in our work. While only two parameter families may conveniently be viewed as a contour plot, the Betti number can be computed for any given set of parameter values. For the topological features to be meaningful for one set of parameter values to the next, there should be a filtration or continuity in the data with changed parameter values. We will demonstrate that even though our data is not a bi-filtration---as there is not inclusion in the simplicial complex from one time step to the next---the simple crocker representation effectively captures topological changes of a time-varying system.

\section*{Results}

We simulate aphid movement using both the interactive and control models, initialized with the same configuration as each of the nine experiments. We run 100 trials for each model-initial condition pair, for a total of 1800 trials. Each simulation has the same number of frames as the experiment from which its initial conditions were taken, ranging in number between 4605 and 5883. We compare the fitness of the two models of aphid motion with the experimental data using two different approaches: first, via several \emph{order parameters}, and then, via topological data in the form of crockers. 

\subsection*{Comparing Models with Order Parameters}
Biological group dynamics are commonly measured with a variety of order parameters designed to give some sense of the collective behavior.  Examples include polarization, angular momentum, absolute angular momentum, and average distance to nearest neighbor; see, \emph{e.g.}, \cite{VicCziBen1995,DOrChuBer2006,HueAld2008}.

The polarization $P \in [0,1]$ measures the group's global agreement on a heading, that is, to what degree individuals are moving in the same direction. Polarization is given by
\begin{equation}
P = \left| \frac{\Sigma_{i = 1}^N \mathbf{v}_i}{\Sigma_{i = 1}^N |\mathbf{v}_i|}\right|,
\end{equation}
where $\mathbf{v}_i$ is the velocity of the $i^{th}$ insect. When all individuals have the same heading, $P=1$. Angular momentum $M_{ang} \in [0,1]$ measures the degree to which the aggregate behavior of the group is rotational. Angular momentum is
\begin{equation}
M_{ang} = \left| \frac{\Sigma_{i = 1}^N \mathbf{r}_i \times \mathbf{v}_i}{\Sigma_{i = 1}^N |\mathbf{r}_i||\mathbf{v}_i|}\right|,
\end{equation}
where $\mathbf{r}_i$ is the position of the $i^{th}$ individual taken relative to the center of mass of the group. When all individuals rotate perfectly and in the same orientation with respect to the center of mass, $M_{ang}=1$. Absolute angular momentum $M_{abs} \in [0,1]$ is
\begin{equation}
M_{abs} = \left| \frac{\Sigma_{i = 1}^N |\mathbf{r}_i \times \mathbf{v}_i|}{\Sigma_{i = 1}^N |\mathbf{r}_i||\mathbf{v}_i|}\right|.
\end{equation}
This quantity is similar to $M_{ang}$, except that it ignores the orientation of rotation. For example, a  group in which some individuals move in clockwise circles around the center of mass and others move in counterclockwise circles (superposed counterrotating vortices) would have low $M_{ang}$, because the two groups cancel out each other's momentum but would have high $M_{abs}$. The average distance to nearest neighbor $d_a \geq 0$ is
\begin{equation}
d_a = \frac{1}{N} \sum_i \min_{j \neq i} |\mathbf{x}_i - \mathbf{x}_j|.
\end{equation}
This quantity gives some sense of the spatial distribution of the group. Finally, because the aphid models under consideration incorporate stop-start motion, we consider the percentage of aphids moving in each frame. In our simulation data, the algorithm outputs each aphid's motion state. In the experimental data, aphids are determined to be stationary if they move a sufficiently small distance from one frame to the next; we take this distance to be $10^{-4}\ m$. For brevity, in all of the order parameter definitions above, we have suppressed notation indicating the dependence on time, even though we will examine order parameter time series below.

We divide these five order parameters into two different categories. The first three, as mentioned previously, are simply order parameters that are frequently used in studies of collective motion. In advance of examining our data, we hypothesize that some of them may not be relevant to pea aphid motion. For example, because aphids do not appear to adjust their headings to align with their neighbors, we expect polarization may not be significant. Similarly, aphids are not known to produce strongly rotational motion, so we expect that the angular momentum order parameters may not be relevant.

The last two order parameters relate to inputs of the model, or, to restate, use \emph{a priori} knowledge of the model. Because distance to nearest neighbor influences aphid step length and aphid motion state in the interactive model, it is natural to wonder whether aphid spatial distribution and overall level of movement will differ between the two models. Because stop-start motion is a key model ingredient, it is also natural to consider the overall level of movement within the group.

We proceed with order parameter analysis as follows. For a generic order parameter $\phi$, we calculate $\phi$ in every frame of an experimental data set, giving rise to a time series $\phi_{exp}(t_s)$, where $s$ indexes the time step. For each model, we calculate the order parameter time series for 100 simulation runs, giving rise to time series $\phi^j_{int}(t_s)$ for the interactive model and $\phi^j_{con}(t_s)$ for the control model, where $j=1,\ldots,100$ indexes simulation. We then compute the average distance between the experimental and model time series via a Euclidean norm,
\begin{equation}
\label{eq:distancedef}
D_{int,con} \equiv \left< \sqrt{\sum_s [\phi_{exp}(t_s) - \phi^j_{int,con}(t_s)]^2} \right>_j.
\end{equation}

With these distances defined, we perform a statistical analysis. Because we do not expect that either model should perfectly describe the complex biological system, we do not perform statistical tests on the null hypothesis that $D_{int}$ and/or $D_{con}$ are zero. Instead, we are interested in assessing whether one model produces data closer to the experiment than the other model. To this end, we calculate the difference in means $ D = D_{con} - D_{int}$ and construct 95\% confidence intervals $D \pm R_{95\%}$, where $R_{95\%}$ is the radius of the interval. In constructing these intervals, we use a Bonferroni correction with $n=81$ to account for the multiple comparison performed in our results, namely nine measures (five order parameters and four topological quantities, introduced below) on nine experiments. If a confidence interval does not include zero, we conclude that one of the models produces data more faithful to the experiment than the other model.

\begin{table}[!ht]
\begin{adjustwidth}{-2.25in}{0in} 
\centering
\caption{\label{table1}
{\bf Summaries of statistical tests comparing models of aphid motion using order parameters.}}
\newcolumntype{?}[1]{!{\vrule width #1}}
\begin{tabular}{crrrrrr?{0.75mm}rrrr}
\hline
& \multicolumn{2}{c}{$P$} 	& \multicolumn{2}{c}{$M_{ang}$} & \multicolumn{2}{c?{0.75mm}}{$M_{abs}$} & \multicolumn{2}{c}{$d_a$} & \multicolumn{2}{c}{$Mov_\%$}\\  
Exp	        		& $D$					& $R_{95\%}$	& $D$					& $R_{95\%}$	& $D$					& $R_{95\%}$	& $D$					& $R_{95\%}$	& $D$   					& $R_{95\%}$ \\ \thickhline
1			& \cellcolor{green!20} 0.17	& 0.13		& \cellcolor{green!20} 1.50	&  0.16		& \cellcolor{red!20} -2.00		& 0.15 		& \cellcolor{green!20} 0.26	& 0.05		& \cellcolor{green!20} 20.2	& 0.40	\\ \hline
2 			& \cellcolor{red!20} -0.25 		& 0.10		& \cellcolor{red!20} -0.97 		&  0.11		& \cellcolor{red!20} -1.66		& 0.09		& \cellcolor{green!20} 0.25	& 0.02		& \cellcolor{green!20} 0.78	& 0.24	\\ \hline
3 			& \cellcolor{green!20} 1.09	& 0.10		& \cellcolor{green!20} 0.40	&  0.11		& \cellcolor{red!20} -1.71		& 0.10		& \cellcolor{green!20} 0.54	& 0.04		& \cellcolor{green!20}25.1		& 0.26	\\ \hline
4 			&  \cellcolor{green!20} 1.70	& 0.08		&  \cellcolor{red!20} -0.78		&  0.09		&  \cellcolor{red!20} -1.24		& 0.07		&  \cellcolor{green!20} 0.16 	& 0.02		& \cellcolor{green!20} 1.68		& 0.21	\\ \hline
5 			& \cellcolor{red!20} -2.22		& 0.09		& \cellcolor{red!20} -0.63		&  0.09		& \cellcolor{red!20} -1.10		& 0.07		& \cellcolor{green!20} 0.31	& 0.03		& \cellcolor{green!20} 7.68	& 0.24	\\ \hline
6			& \cellcolor{green!20} 1.04 	& 0.09		& \cellcolor{red!20} -0.49		&  0.10		& \cellcolor{red!20} -1.68		& 0.09		& \cellcolor{green!20} 0.53	& 0.04		& \cellcolor{green!20}24.0		& 0.26	\\ \hline
7			& \cellcolor{red!20} -0.97 		& 0.11		& \cellcolor{red!20} -1.28		&  0.13		& \cellcolor{red!20} -1.99		& 0.11		& \cellcolor{green!20} 0.57	& 0.04		& \cellcolor{green!20} 23.9	& 0.29	\\ \hline
8			& \cellcolor{red!20} -3.30		& 0.25		& \cellcolor{red!20} -2.11		&  0.22		& \cellcolor{red!20} -2.87		& 0.26		& \cellcolor{green!20} 0.57	& 0.25		& \cellcolor{green!20} 13.2	& 0.63	\\ \hline
9			& \cellcolor{red!20} -4.69		& 0.25		& \cellcolor{red!20} -3.14		&  0.22		& \cellcolor{red!20} -2.81		& 0.20		& \cellcolor{green!20} 0.48	& 0.15		& \cellcolor{green!20} 11.7		& 0.34	\\ \hline
\end{tabular}
\begin{flushleft} Data to the left of the thick vertical line summarize statistical tests performed on three order parameters that do not use \emph{a priori} knowledge of the models: polarization $P$, angular momentum $M_{ang}$, and absolute angular momentum $M_{abs}$. Data to the right of the thick vertical line correspond to two order parameters that do use \emph{a priori} knowledge of the models:  average distance to nearest neighbor $d_a$ and overall percentage of insects moving $Mov_{\%}$. For each order parameter and each experiment, we calculate the average Euclidean distance $D_{int}$ between the experimental data and data produced by 100 runs of the interactive model seeded with the same initial condition as the experiment. We also calculate $D_{con}$, which is similar, but for the control model. See Equation (\ref{eq:distancedef}) and main text for more detail. To assess whether one model produces data closer to the experiment than the other model, we calculate $D = D_{con} - D_{int}$, shown in the table. The columns $R_{95\%}$ provide Bonferroni-corrected 95\% confidence radii on $D$. If the interval $[D-R_{95\%},D+R_{95\%}]$ does not include zero, we conclude that one model produces data more faithful to the experiment than the other model. Cases where the confidence interval excludes zero are colored green or red depending on whether the interactive model or control model, respectively, is more faithful to experiment.
\end{flushleft}
\end{adjustwidth}
\end{table}

Table~\ref{table1} summarizes results of our statistical tests on order parameters. Data to the left of the thick vertical line summarize statistical tests performed on the three order parameters that do not use \emph{a priori} knowledge of the models: polarization $P$, angular momentum $M_{ang}$, and absolute angular momentum $M_{abs}$. Data to the right of the thick vertical line correspond to the two order parameters that do use \emph{a priori} knowledge of the models: average distance to nearest neighbor $d_a$ and overall percentage of insects moving $Mov_{\%}$. Even with Bonferroni corrections, all statistical comparisons show up as significant. We color results green or red depending on whether the interactive model or control model, respectively, is more faithful to experiment.

The order parameters that do not use \emph{a priori} knowledge of the models give a mixed message. For polarization $P$, the interactive model (green) agrees better with experimental data for four of the experimental trials and the control model (red) agrees better for the other five. Angular momentum $M_{ang}$ is split as well; we see the control model as a closer fit in seven out of nine trials. Finally, for the absolute angular momentum $M_{abs}$, the control model is closer for all trials. We conjecture that this somewhat counterintuitive result is related to the reflective boundary condition implemented in both models, which may impact the model results differentially and may be a poor reflection of biological reality.

In contrast, the order parameters that do use \emph{a priori} knowledge of the models give a consistent message. For both average distance to nearest neighbor $d_a$ and percentage of aphids moving $Mov_\%$, the interactive model data is significantly closer to the experimental data than the control model is. This is not surprising as these order parameters capture properties closely related to model inputs hypothesized to be important during the modeling process, namely the location of the nearest neighbor and motion state transitions.

\subsection*{Comparing Models with Topology}

Momentarily, we will present results comparing the two aphid motion models with experiment using topology. First, to aid this presentation, we further discuss how to interpret crockers to track aphid movement. 

\begin{figure}[!h]
\centerline{\includegraphics[width=0.9\textwidth]{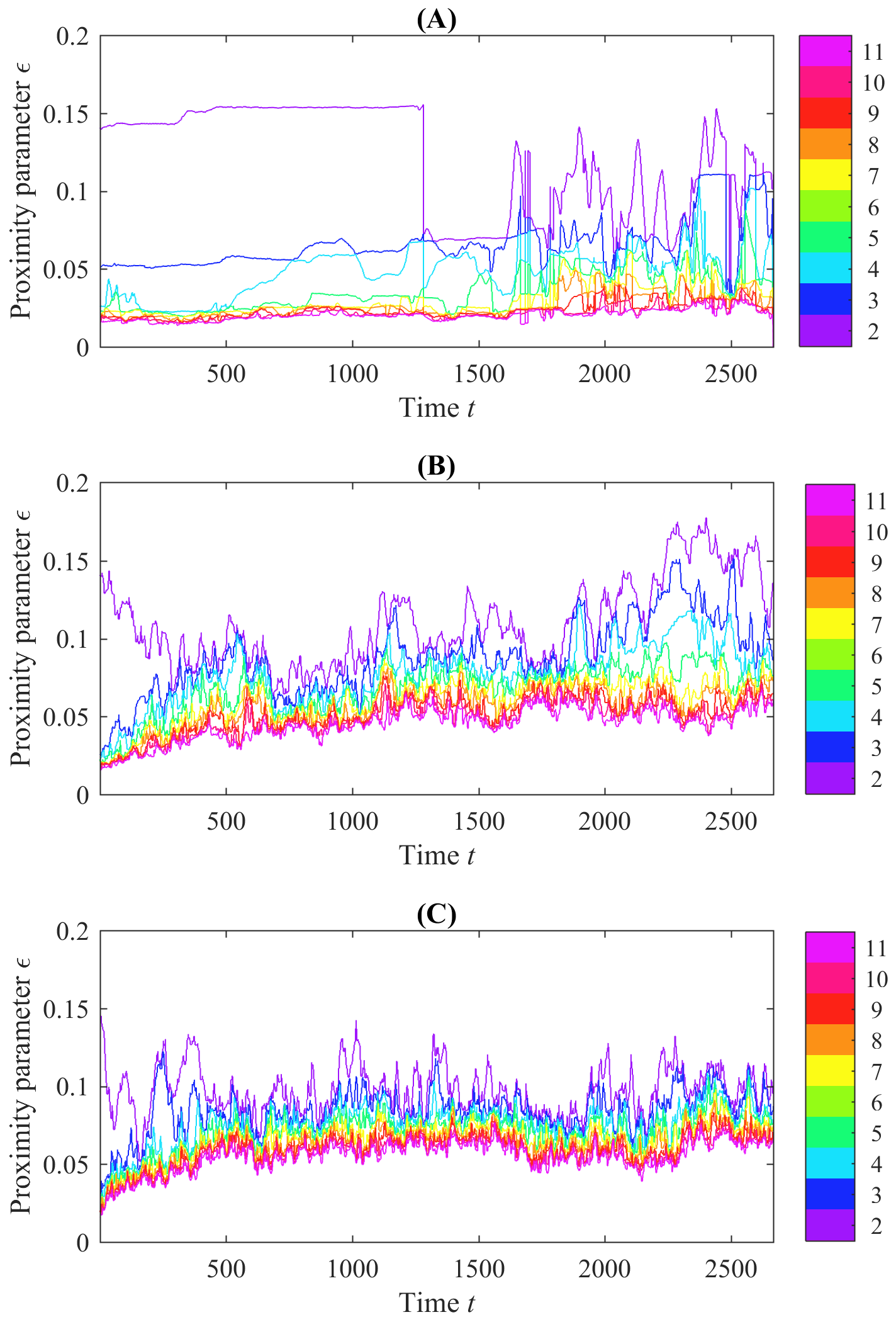}}
\caption{{\bf Crocker plots of experimental and simulated position data.}
(A) shows the crocker plot of the experimental data from Experiment 7. (B) and (C) show crocker plots from sample realizations of (respectively) the interactive model and control model, seeded with initial conditions from Experiment 7. At $t=0$, the highest contour separate from the rest represents a lone aphid; it takes high $\epsilon$ values to connect it to the others. Looking at the initial conditions for Experiment 7 shown in Fig \ref{fig1}, we can see the isolated aphid at the lower right.}
\label{fig5}
\end{figure}

Fig~\ref{fig5}(A) shows the $\betti_0$ crocker plot corresponding to the position data from Experiment 7. Figs~\ref{fig5}(B) and \ref{fig5}(C) show analogous information for realizations of the interactive and control models, seeded with the same initial condition as in Fig~\ref{fig5}(A). Because these plots show $\betti_0$ and are based on position data, we refer to them as  $\betti_{0(pos)}$ crocker plots. Each crocker plot starts with many connected components for small values of $\epsilon$, and the number of components shrinks as $\epsilon$ increases and components are joined. We display only contours of 11 or less, inferring higher contours as topological noise. Since each of the trials begins with aphids in the same place, we would expect the contours in each crocker plot to begin at the same levels. This is the case for most contours. However, the blue contour, which represents the boundary between the regions with two and three connected components, appears to begin higher in the experimental crocker plot than it does in the two simulated crocker plots. This is due to a single aphid being lost by the motion-tracking software for one frame, right after the start of the experiment. Motion-tracking software is imperfect, and in the original experimental data, it is common for aphids to be dropped and then picked back up in later frames. This phenomenon is also responsible for the abrupt vertical changes present in some of the contours in the experimental crocker plots.

In all three panels of Fig~\ref{fig5}, there is an entity which is isolated in early frames and only connects to the rest of the aphids for larger values of $\epsilon$. Looking at the initial conditions for Experiment 7, shown in Fig~\ref{fig1}, it is clear that this entity is the single aphid at the bottom-right of the arena. In the crocker plot representing the experimental data, this aphid does not move closer to the others until nearly halfway through the trial, though the vertical drop suggests that at this point it was lost by the motion-tracking software. However, in both of the simulation trials, the aphid joins the others earlier, a result of the models' randomness and tendency toward aggregation.

\begin{figure}[!h]
\centerline{\includegraphics[width=0.9\textwidth]{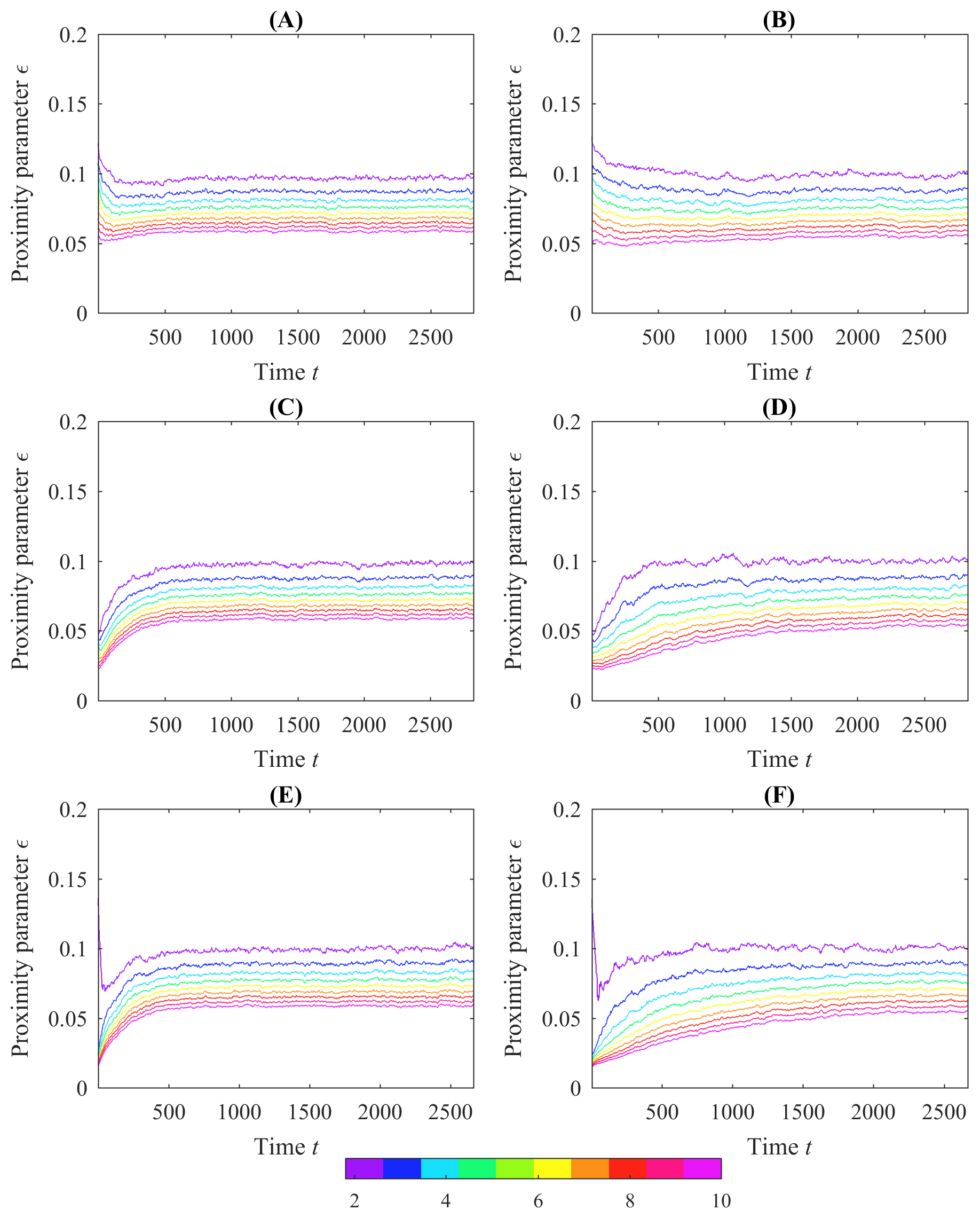}}
\caption{{\bf Examples of average $\betti_{0(pos)}$ crocker plots over 100 runs of the two models simulated using initial conditions from experiments.}
The left column corresponds to position data simulated with the control model and the right from the interactive model. The first row corresponds to Experiment 2, the second from Experiment 3, and the third from Experiment 7. In general, crocker plots arising from experiments with more tightly clustered initial configurations exhibit upward trending of the contours over time in contrast to experiments where aphids are more dispersed initially; refer to Fig \ref{fig1}.}
\label{fig6}
\end{figure}

Additionally, both of the sample simulation trials show an upward trend in the crocker contours over time. This trend is even more pronounced in the average $\betti_{0(pos)}$ crocker plots for each of the models, computed by taking the element-wise average of the topological data over 100 simulations. This averaging reduces noise as illustrated in Fig \ref{fig6}. The upward trend in the contours signifies that aphids are moving farther apart, dispersing in the arena. That is, when the contours are low, small values of $\epsilon$ serve to connect all the aphids into connected components; when the contours are higher, connecting the aphids requires larger values of $\epsilon$, meaning the aphids themselves are farther apart. 

This upward trend is present in the crocker plots of simulation trials initialized with the configurations from Experiments 1, 3, 6, 7, 8, and 9, and to a lesser extent in those with initial conditions from Experiment 4. In simulation trials with initial conditions from Experiments 2 and 5, however, the contours of the crocker plots are level or even have a slight initial dip. Average $\betti_{0(pos)}$ crocker plots for Experiments 2, 3, and 7 are shown in Fig~\ref{fig6} as examples. In contrast, these same patterns are not apparent in the crocker plots of the experimental data. Looking at the initial conditions in Fig~\ref{fig1}, we see that crocker plots that exhibit upward-trending contours correspond to initial conditions where aphids are clustered tightly; the plots without upward-trending contours are from initial conditions in which the aphids are initially more dispersed. This suggests that in both models, aphids are dispersing (or, in the case of simulation trials with initial conditions from Experiments 2 and 5, aggregating) toward a preferred social distance. Additionally, the average plots from the control model rise more quickly than plots from the interactive model. Since these trends are not present in the experimental crocker plots, this preferred distance does not appear to be present in real aphid interactions. This suggests that both models could be improved to better reflect patterns of movement present in the experimental data. It is notable that our observation of this trending phenomenon emerged from the crocker plot representation, was not apparent in the previous analysis of these models \cite{NilPaiWar2013}, and was not a known characteristic \emph{a priori}.

We now turn to a discussion of using topology to compare the fitness of the two models with experiment. We consider two sets of data, just aphid position (2-dimensional) and (normalized) aphid position and velocity concatenated (4-dimensional). In order to fairly compare position and velocity data simultaneously in the second set of data, we normalize both position and velocity, so they are on the same scale. According to \cite{NilPaiWar2013}, the longest aphid step length is $0.0013\ m,$ and the arena has radius $0.2\ m$. Therefore, we normalize the position data by 0.2 and the velocity by 0.0013. We compute persistent homology of the 2-D and 4-D point clouds at each time step (downsampled by a factor of four to speed up computations) to generate interval data corresponding to topological features using the maximum filtration value of $\epsilon=0.2$ for the position data and $\epsilon=1.5$ for the concatenated position and velocity data. We convert the interval data to crocker plots corresponding to dimensions $k=0, 1$ (denoted as $\betti_{0(pos)}$, $\betti_{1(pos)}$, $\betti_{0(posvel)}$, $\betti_{1(posvel)}$, respectively),  uniformly sampling 50 proximity values from 0 up to the maximum filtration. 

We proceed with the topological analysis in the same manner as for the order parameters. Given the matrix corresponding to a crocker plot $C$, we compare the experimental crocker plot $C_{exp}$ to each of the 100 simulations runs of both the interactive and control models, denoted $C_{int}^j$ and $C_{con}^j$, respectively, where $j=1, \ldots, 100$ indexes simulation. We compute the average distance between the crocker plots of the experiment and the two models by taking the Frobenius norm of the matrices corresponding to the crocker plots
\begin{equation}
\label{eq:distancedefcrocker}
D_{int,con} \equiv \left< ||C_{exp} - C^j_{int,con}||_{Fro} \right>_j
\end{equation}
As before, we compute the difference in means $D = D_{con} - D_{int}$ and the 95\% confidence radii on $D$. 

Table \ref{table2} summarizes results of our statistical tests on the topological data. Again, even with Bonferroni corrections, all statistical comparisons show up as significant. We color results green or red depending on whether the interactive model or control model, respectively, is more faithful to experiment. All four of the topological representations give a consistent message that the interactive model crockers are significantly closer to the experimental crockers than the control model. The consistency revealed by these statistical tests shows that the topological crocker representation captures characteristics of the global motion of aphids that agrees with intuition behind the model development, namely that the model accounting for social interactions is more consistent with the experiment. The key point is that the topological approach uses no \emph{a priori} knowledge about the data sets, nor the models that generated them.

\begin{table}[!ht]
\begin{adjustwidth}{-2.25in}{0in} 
\centering
\caption{\label{table2}
{\bf Summaries of statistical tests comparing models of aphid motion using topology.}}
\newcolumntype{?}[1]{!{\vrule width #1}}
\begin{tabular}{crrrrrrrr}
\hline
& \multicolumn{2}{c}{$\betti_{0(pos)}$} & \multicolumn{2}{c}{$\betti_{1(pos)}$}& \multicolumn{2}{c}{$\betti_{0(posvel)}$} & \multicolumn{2}{c}{$\betti_{1(posvel)}$}\\
Exp & $D^*$ & $R_{95\%}^*$ & $D^*$ & $R_{95\%}^*$ & $D^*$ & $R_{95\%}^*$ & $D^*$ & $R_{95\%}^*$ \\ \thickhline
1 & \cellcolor{green!20}  1147 & 38.73 & \cellcolor{green!20} 135.1 & 5.149 & \cellcolor{green!20} 831.9 & 42.81& \cellcolor{green!20} 124.2 & 5.640\\ \hline
2 & \cellcolor{green!20} 1703 & 23.98 & \cellcolor{green!20} 297.0 & 6.500 & \cellcolor{green!20} 858.9 & 16.88& \cellcolor{green!20} 261.1 & 7.258\\ \hline
3 & \cellcolor{green!20} 1893 & 37.36 & \cellcolor{green!20} 290.4 & 6.836 & \cellcolor{green!20} 1347 & 35.60&\cellcolor{green!20} 271.3 & 7.218\\ \hline
4 & \cellcolor{green!20} 1825 & 18.36 & \cellcolor{green!20} 359.0 & 5.741 & \cellcolor{green!20} 552.3 & 19.04& \cellcolor{green!20} 332.5 & 6.301\\ \hline
5 & \cellcolor{green!20} 1370 & 15.81 & \cellcolor{green!20} 249.2 & 6.039& \cellcolor{green!20} 270.8 & 15.25& \cellcolor{green!20} 225.1 & 6.401\\ \hline
6 & \cellcolor{green!20} 2086 & 40.53 & \cellcolor{green!20} 322.9 & 7.061 & \cellcolor{green!20} 1484 & 37.35& \cellcolor{green!20} 307.7 & 7.116\\ \hline
7 & \cellcolor{green!20} 1747 & 34.32 & \cellcolor{green!20} 264.8 & 6.235& \cellcolor{green!20} 1289 & 33.69& \cellcolor{green!20} 247.5 & 6.69\\ \hline
8 & \cellcolor{green!20} 298.5 & 20.81 & \cellcolor{green!20} 10.28 & 1.920& \cellcolor{green!20} 147.9 & 19.81& \cellcolor{green!20} 3.573 & 1.641\\ \hline
9 &\cellcolor{green!20} 467.5 & 19.66 & \cellcolor{green!20} 22.34 & 2.148 & \cellcolor{green!20}245.3 & 21.10& \cellcolor{green!20} 10.93 & 2.067\\ \hline
\end{tabular}
\begin{flushleft} These data summarize statistical tests based on four choices of topological measures: the zeroth and first Betti numbers of aphid position data as well as (normalized) aphid position and velocity concatenated, $\betti_{0(pos)}$, $\betti_{1(pos)}$, $\betti_{0(posvel)}$, and $\betti_{1(posvel)}$. Results are analogous to Table~\ref{table1} but rather than using a Euclidean norm of the differences of vectors representing order parameter time series, we use the Frobenius norm of the differences of matrices representing crocker plots. See Equation (\ref{eq:distancedefcrocker}) and main text for more detail. As in Table~\ref{table1}, cases where the Bonferroni-corrected 95\% confidence radii on $D = D_{con} - D_{int}$ confidence interval, $R_{95\%}$, exclude zero are colored green or red depending on whether the interactive model or control model, respectively, is more faithful to experiment.
\end{flushleft}
\end{adjustwidth}
\end{table} 

In addition to these results, we perform analogous statistical tests on concatenated $\betti_{0(pos)}$ and $\betti_{1(pos)}$ crockers as well as concatenated $\betti_{0(posvel)}$ and $\betti_{1(posvel)}$ crockers. The results are similar to the dimension $k=0$ representations in Table \ref{table2} as the $k=1$ features are sparse. 

\section*{Discussion and Conclusion}

We have assessed the goodness-of-fit of two models of aphid motion: an interactive model that describes social behavior of the insects and a control model that omits this effect. To compare the model results to experimental data, we performed statistical tests on three different sets of measures: order parameters that do not use \emph{a priori} knowledge of the models, order parameters that do use this information, and topological measures in the form of crocker plots. Order parameters that rely on \emph{a priori} knowledge give a consistent message, namely that the interactive model better describes the experimental data. The topological measures match the performance of the \emph{a priori} order parameters and outperform the other order parameters. Of course, the topological measures do not rely on knowledge of the models. Thus, we find that adopting a topological lens is a useful approach for characterizing and comparing motion of biological groups when one has little information about what model mechanisms might be important.

Crocker representations have the potential to encode richer information than traditional order parameters. Consider aphid proximity. The configuration of any aphids at a fixed moment in time will be joined into a single connected component when $\epsilon$ reaches the maximum distance between any pair of aphids. On the other hand, the nearest neighbor order parameter condenses information into an average, losing nuance that the contours in the crocker plot retain. As a concrete example, consider the following two scenarios. In the first, imagine a single cluster of 10 objects where the average distance to an object's nearest neighbor is $2\ cm$. In the second, imagine ten objects  scattered in five pairs, where the two objects within each pair are $2\ cm$ apart, but two aphids from different pairs are at least $5\ cm$ from each other. The nearest neighbor order parameter would be equal to 2 in both cases but the $\betti_0$ crocker plots would be able to distinguish between differing configurations. In the crocker for the first case, there would be only one component for $\epsilon \geq 2$. However, in the second crocker, there would be five connected components at $\epsilon = 2$, and the number would only decrease for $\epsilon \geq 5$. Thus, while the average distance to nearest neighbor order parameter and the crocker plot are related, the latter contains more information.

One limit of our approach involves computational complexity. Computation of order parameters involves determining a single number for each time step. Crockers, on the other hand, compute persistent homology at each time step, which is a multi-scale approach and is thus a more costly computation. Developing fast software to compute persistence is an active area of research in the topological data analysis community, and recent advances have sped up computations by several orders of magnitude; see \cite{OttPorTill2017} for benchmarking on a set of algorithms and implementations. Nevertheless, computing persistent homology will be more computationally expensive than computing an order parameter. Still, we have found it to reveal useful information about the structure of the data.

Another limit of our approach involves our choice of norm. We have selected the Euclidean norm to compute the distance between order parameters. Therefore, for consistency, we selected the Frobenius norm to compute the distance between crocker matrices, since it is analogous to the Euclidean norm on a vectorized matrix. Another natural choice of metric on crockers might be the Wasserstein distance, commonly called the earth mover's distance. Selecting other distance measures could change our results, and we leave this as future work.

\section*{Acknowledgments}
MU was supported by a Macalester College Wallace Scholarly Activities grant given to LZ. CT is supported by National Science Foundation grant DMS-1813752. We are grateful to Andrew Bernoff and Devin Bjelland for helpful conversations.

\nolinenumbers

%
%


\end{document}